\documentclass[a4paper,fleqn,usenatbib]{mnras}
\usepackage{mathptmx}
\usepackage[T1]{fontenc}
\usepackage{ae,aecompl}
\usepackage{multirow,color}
\usepackage{dcolumn}
\usepackage{bm}
\usepackage{amssymb,amsmath,amsfonts,graphicx,epsf}
\usepackage{color}
\usepackage{enumerate}

\newcommand{\bvec}[1]{\mathbf{#1}}
\newcommand{\pa}{\partial}

\title{Electron Strahl and Halo Formation in the Solar Wind}

\author[K. Horaites et al.]{
Konstantinos Horaites,$^{1}$\thanks{E-mail: horaites@wisc.edu}
Stanislav Boldyrev,$^{1,2}$
Mikhail V. Medvedev$^{3,4}$
\\
$^{1}$Department of Physics, University of Wisconsin -- Madison, 1150 University Avenue, Madison, WI 53706, USA\\
$^{2}$Space Science Institute, Boulder, CO 80301, USA\\
$^{3}$Department of Physics and Astronomy, University of Kansas, Lawrence, KS 66045\\
$^{4}$Laboratory for Nuclear Science, Massachusetts Institute of Technology, Cambridge, MA 02139
}

\date{Accepted XXX. Received YYY; in original form ZZZ}

\pubyear{2018}

\begin{document}
\label{firstpage}
\pagerange{\pageref{firstpage}--\pageref{lastpage}}
\maketitle

\begin{abstract}
We propose a kinetic model describing the formation of the strahl and halo electron populations in the solar wind. We demonstrate that the suprathermal electrons propagating from the sun along the Parker-spiral magnetic field lines are progressively focused into a narrow strahl at heliospheric distances $r\lesssim 1$~AU, while at $r\gtrsim 1$~AU the width of the strahl saturates due to Coulomb collisions and becomes independent of the distance. Our theory of the strahl broadening does not contain free parameters and it agrees with the Wind observations at~1~AU to within~$15-20\%$. This indicates that Coulomb scattering, rather than anomalous turbulent diffusion, plays a dominant role in strahl formation in these observations.  We further propose that the nearly isotropic halo electron population may be composed of electrons that ran away from the sun as an electron strahl, but later ended up on magnetic field lines leading them back to the sun. Through the effects of magnetic defocusing and Coulomb pitch-angle scattering, a narrow source distribution at large heliocentric distances appears nearly isotropic at distances $\sim$1~AU. The halo electrons are, therefore, not produced locally; rather, they are the fast electrons trapped by magnetic field lines on global heliospheric scales.  At the electron energies $K \lesssim 200\,\, {\rm eV}$, our theory is quite insensitive to the particular mechanism that produces a population of sunward-streaming suprathermal particles.  At larger energies, however, our theory indicates that the scattering provided by Coulomb collisions alone is not sufficient to isotropize a narrow sunward-propagating electron beam. 
\end{abstract}


\maketitle

\section{Introduction}
Electrons in the solar wind are only weakly regulated by Coulomb collisions. The electron velocity distribution functions (eVDFs) exhibit features in addition to the Maxwellian ``core'' of the distribution, which comprises the bulk of the density. At higher energies (e.g., between approximately $10\,\,{\rm eV}$ and $1\,\,{\rm keV}$ at 1~AU) the distribution exhibits a field-aligned beam known as the ``strahl'', and a nearly-isotropic component known as the ``halo'' \citep[e.g.,][]{feldman75,pilipp87}. Due to their relatively high energies, the strahl and halo electron populations are less affected by Coulomb collisions than the electron core, and particles in these populations can travel over heliospheric scales without coming into thermal equilibrium with the ambient plasma~\cite[e.g.,][]{scudderolbert79}.  

The strahl is believed to represent runaway electrons originating in the hot regions of the inner heliosphere ($\sim~5-15~\,r_{\odot}$). The beam-like shape of this population results from the competition of two kinetic effects: the beam focuses as particles propagate into a spatially weakening magnetic field, but is broadened as a result of electron pitch-angle scattering. The pitch-angle scattering can be provided by Coulomb collisions, but additional sources---for example, electron interactions with plasma turbulence---may contribute to so-called ``anomalous diffusion'' of the distribution. Whistler-mode turbulence has been identified (both theoretically and observationally) as a potential source of anomalous scattering; such turbulence may pre-exist in the plasma or may be generated by the highly anisotropic electron velocity distribution itself \cite[e.g.,][]{gary75, gary94, vocks03,vocks05,saitogary07,pagel07, wilson13, lacombe14, stansby16}. The relative contributions of Coulomb scattering and anomalous scattering have not been well constrained by the existing observations. While the observations gravitate to the conclusion that the width of the electron strahl is typically larger than the bare minimum predicted by classical Coulomb collisions, significant evidence shows the properties of the electron strahl are correlated to the degree of Coulomb collisionality between the strahl particles and the core population \cite[][]{feldman78,scudderolbert79,lemonsfeldman83,pilipp87,ogilvie00,salem03,anderson12,bale13, horaites18a}.        

The halo component of the eVDF is the least well understood theoretically, in spite of the fact that the halo is observed ubiquitously in the solar wind and its properties have been thoroughly characterized by observations. A reasonable theory of the halo needs to explain a variety of its characteristics, most notably, its high degree of isotropy and its non-Maxwellian velocity profile.  
Several competing theories have been posited over the years that have attempted to explain the halo's origin. \cite{scudderolbert79} proposed that the sunward-moving suprathermal electrons observed near 1 AU---which are generally classified as part of the halo population---had previously been moving anti-sunward but had undergone large-angle backscattering at some larger heliocentric distance 1-10 AU. \cite{vocks05} developed numerical simulations in which a spectrum of sunward-propagating whistler waves was imposed. These waves supplied angular diffusion, in concert with Coulomb collisions, that developed the suprathermal electrons into halo and strahl populations. \cite{che14}, and \cite{chegoldstein14} noted that the two-stream instability caused by very energetic electrons ($\sim 1\,\,{\rm keV}$) may play a role in establishing the relative magnitudes of the strahl and halo distributions. \cite{livadiotismccomas11} applied non-extensive statistical mechanics to address the non-Gaussian shape of the halo velocity distribution function. \cite{lichko17} proposed that the halo electrons may be accelerated to suprathermal energies by the magnetic pumping mechanism, which arises from the repeated large-scale compression and decompression of magnetic field lines.

In our previous work \cite[][]{horaites18a}, we developed a theory for the electron strahl based on Coulomb collisions. We demonstrated that the drift-kinetic equation for the fast strahl electrons can be solved if the variations of the magnetic field and plasma density are known as functions of distance along a magnetic flux tube. We approximated these dependencies from local measurements of the plasma parameters, assuming that the background density, temperature, and magnetic field strength all varied as power laws of distance along the flux tube. We found that our derived expression for the strahl width provided a surprisingly good fit to Wind observations of the eVDF at 1 AU, even though only Coulomb scattering was taken into account. In \cite{horaites18b}, we discussed the linear kinetic stability of a simple core-strahl eVDF; surprisingly, our numerical analysis did not reveal any unstable modes that would resonate with the fast-moving strahl particles.

In the present work, we further generalize our theory of the electron strahl velocity distribution function, assuming only that the global magnetic field follows the azimuthally symmetric Parker spiral. Our prediction for the strahl width does not contain free parameters and is applicable for an arbitrary heliospheric distance.\footnote{To be more precise, in our treatment of fast electrons ($10-100$~eV) we neglect the time-variation of the magnetic-line structure, since the speed of the electrons is much larger that that of the solar wind. This assumption, however, becomes incorrect at very large heliospheric distances, where the Parker-spiral magnetic field is mostly azimuthal and the {\em radial} velocity of the electrons streaming along the magnetic field lines becomes comparable to the solar wind velocity. At those distances ($r\gtrsim 20$~AU), our assumption of stationarity does not hold, and our theory needs to be modified to include the solar-wind advection effects.} We find that while the runaway electron beam is focused by the radially weakening magnetic field, its width {\em saturates} and approaches a universal value at distances beyond~$1\,{\rm AU}$. For the strahl particles of energy $100\,{\rm eV}$ in a plasma of ambient density $5 \,{\rm cm}^{-3}$, the total strahl width saturates at about $24^\circ$ irrespective of the distance as long as $r\gg 1\,{\rm AU}$. The magnetic-field focusing effects are thus effectively arrested by classical Coulomb collisions in the outer heliosphere, yielding a significant saturated strahl width.

The saturation of the strahl-width variation with distance is an important result of our Coulomb theory.  
We therefore suggest that when the observations find that the strahl width does not saturate but rather increases with the distance at $r>1$~AU, see, e.g, the discussion in \cite{anderson12,graham17,graham18}, some other scattering mechanisms, in addition to Coulomb collisions, must be at play.  However, a direct comparison of our prediction with the measurements of Wind satellite at 1~AU shows that our Coulomb theory describes the strahl width rather well, leaving limited room for anomalous scattering effects in the region $r \lesssim 1$ AU.

We further propose that if the strahl particles collimated in this process later find their way back to the sun, say, by following closed magnetic field lines \cite[e.g.,][]{gosling1993,gosling2001}, then at least for the particle energies of $K\lesssim 200\,\,{\rm eV}$, their velocity distribution will inevitably approach an isotropic shape at lower heliospheric distances due to the combined effect of magnetic defocusing and pitch-angle Coulomb scattering. Such isotropization is very efficient; for it to occur the strahl electrons with an energy of~$100\,\,{\rm eV}$ need to be turned around at a distance of about~$8\,{\rm AU}$.  The halo electrons in our model are, therefore, not produced locally; rather, they are the runaway electrons trapped by magnetic field lines on a global heliospheric scale.  Our theory is, however, insensitive to the particular mechanism that produces a population of sunward-streaming suprathermal particles at large distances. As long as such a population exists, it should develop naturally into an isotropic distribution at smaller distances. Our analysis demonstrates that at least in the energy range $K\lesssim 200\,\,{\rm eV}$ classical Coulomb collisions can play a dominant role in producing the halo population, thus circumventing the need to invoke strong anomalous diffusion mechanisms (e.g., strong wave-particle scattering). At higher electron energies, however, our theory demonstrates that Coulomb collisions become significantly less effective, and they alone cannot isotropize a narrow electron beam. 

In Section \ref{strahl_sec}, we derive an analytic expression for the strahl distribution, that uniquely predicts this population's angular width. In Section \ref{obs_sec}, we compare the prediction for the strahl width with direct measurements derived from the Wind satellite's SWE strahl detector \citep{ogilvie95}. In Section \ref{halo_sec}, we present a speculative theory of the halo distribution, that like our strahl theory, is based only on magnetic focusing and Coulomb collisions. Our conclusions are presented in Section \ref{conclusions_sec}.

\section{The electron strahl}\label{strahl_sec}

We will describe the electron distribution function $f$ in terms of the distance along a magnetic flux tube $x$, the velocity magnitude $v$, and cosine of the pitch angle $\mu$:
\begin{eqnarray}
 \mu\equiv \hat B \cdot \vec{v}/v,
\end{eqnarray}
where the unit vector $\hat B$ points along the (Parker spiral) magnetic field, in the anti-sunward orientation. The steady-state drift kinetic equation for the distribution function $f(v, \mu, x)$  then takes the following form \citep[e.g.,][]{kulsrud83}: 
\begin{eqnarray}
\label{drift_kinetic_eq_udzero}
\mu v \frac{\pa f}{\pa x} &-& \frac{1}{2}\frac{d\ln B}{dx}v (1-\mu^2) \frac{\pa f}{\pa \mu} - \nonumber \\
& -&\frac{e E_\parallel}{m} \left[\frac{1-\mu^2}{v} \frac{\pa f}{\pa \mu}  + \mu \frac{\pa f}{\pa v}\right] = \hat C(f),
\end{eqnarray}
where $E_\|$ is the electric field parallel to the magnetic field line. In Equation (\ref{drift_kinetic_eq_udzero}) we have neglected the $\bvec{E} \times \bvec{B}$ drift. This equation describes the evolution of the (gyrotropic) electron distribution function for the electron population whose speed is much greater than the speed of the solar wind, $v\gg v_{sw}$. The magnetic fields lines are advected with the solar wind, and, therefore, the magnetic field can be assumed stationary for such electrons. If we further assume that the energies of these electrons significantly exceed the thermal energy of the core particles, we may use the linearized form of the collision integral for such electrons \cite[e.g.,][]{helandersigmar02}:
\begin{eqnarray}\label{coll_op_eq}
\hat C(f)  = \frac{4 \pi n e^4 \Lambda}{m_e^2} \left[\frac{\beta}{v^3}\frac{\pa}{\pa \mu} (1-\mu^2)\frac{\pa f}{\pa \mu}  \right. \nonumber \\
   +\left. \frac{1}{v^2}\frac{\pa f}{\pa v} + \frac{v_{th}^2}{2v^2}\left( -\frac{1}{v^2}\frac{\pa f}{\pa v} + \frac{1}{v}\frac{\pa^2 f}{\pa v^2}\right) \right] ,
\end{eqnarray}
where $\Lambda$ is the Coulomb logarithm, $\beta=(1+Z_{eff})/2$, $Z_{eff}=\sum_i n_iZ_i^2/n_e$ is the effective ion charge, and $n \approx n_e$ and $v_{th}$ are respectively the density and thermal speed of the core electron population. As is typical for the fast solar wind $v_{sw} > 550$ km/sec, we will assume that the abundance of ${\rm He}^{2+}$ is 5\% of the ${\rm H}^+$ abundance \citep{wurz05}, and neglect the minor ions. For a quasi-neutral plasma with this composition, we find $Z_{eff}\approx 1.1$ and $\beta \approx 1.05$. The first term in the collision integral~(\ref{coll_op_eq}) describes the pitch-angle scattering of the fast electrons by the slow ions and electrons of the core population, while the remaining terms describe the energy exchange with the core electrons. If we are interested in the evolution of the fast electrons forming a narrow electron strahl with $v_\|\gg v_\perp$, one can demonstrate that the energy-exchange term is negligible in comparison to the scattering term. In what follows, we therefore keep only the first term in Eq.~(\ref{coll_op_eq}). 

We simplify the analysis by introducing new variables, proportional to the electron energy, $E=v^2+(2/m_e)e\phi(x)$, and the magnetic moment, $M=(1-\mu^2)v^2/B(x)$. In these expressions, $e<0$ is the electron charge, and $\phi(x)$ is the electric potential measured with respect to $x=\infty$. In these variables, the drift-kinetic equation (\ref{drift_kinetic_eq_udzero}) for the electron distribution function $f(E, M, x)$ takes a simple form:
\begin{eqnarray}
\frac{\partial f}{\partial x} =\frac{16\pi e^4 \Lambda \beta n(x)}{{\cal E}(E,x)B(x)}\frac{\partial }{\partial M}M\sqrt{1-\frac{MB(x)}{{\cal E}(E,x)}} \frac{\partial f}{\partial M},\label{diffusion}
\end{eqnarray}
where ${\cal E}(E, x)=E-(2/m_e)e\phi(x)$. For the runaway strahl electrons that we consider, ${\cal E}\approx E$, and since $v_\|\gg v_\perp$ we have $MB(x)/{\cal E}(E,x)\ll 1$. Equation~(\ref{diffusion}) can then be simplified as:
\begin{eqnarray}
\frac{\partial f}{\partial x} =\frac{16\pi e^4 \Lambda \beta n(x)}{{E}B(x)}\frac{\partial }{\partial M}M \frac{\partial f}{\partial M}.\label{drift-diffusion}
\end{eqnarray}
We can now introduce a new spatial variable $y$ from the condition
\begin{eqnarray}
dy=\left(\frac{16\pi e^4\Lambda \beta}{E} \right)\left(\frac{n(x)}{B(x)} \right)\,dx,\label{dy}
\end{eqnarray}
where $dx$ is the length element along the magnetic field line. We now notice that the vector ${\bf B}(x)/n(x)$ is frozen into the plasma flow. This means that:
\begin{eqnarray}
dx/dx_0=\left(\frac{B(x)}{n(x)} \right)/\left(\frac{B(x_0)}{n(x_0)} \right),\label{frozen}
\end{eqnarray} 
where $x_0$ is some fixed position, which we will choose, for definiteness, as the point where the magnetic field line is directed at $45^\circ$ with respect to the radial direction. For simplicity we will assume an axisymmetric Parker spiral model for the magnetic field. For the Parker spiral, the heliospheric distance $r_0$ corresponding to the point $x_0$ is given by the formula:
\begin{eqnarray}\label{r_0_eq}
r_0=v_{sw}/\omega_s,
\end{eqnarray}
where $\omega_s$ is the model angular velocity of the sun and $v_{sw}$ is the (constant) speed of the solar wind. In practice, the heliospheric distance $r_0$~corresponding to the point $x_0$ turns out to be approximately~1~AU.
\begin{figure}
\includegraphics[width=1\linewidth]{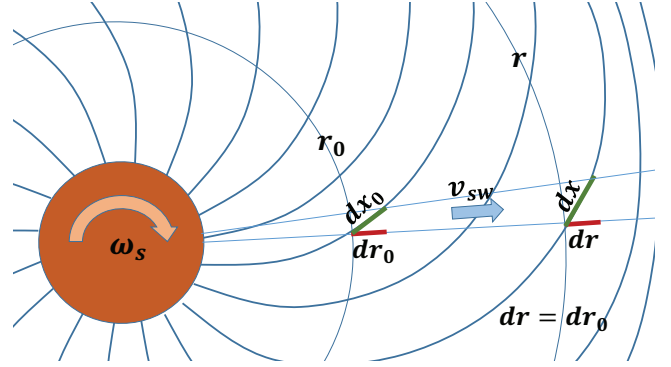}
\caption{Sketch (drawn not to scale) of the magnetic field lines forming an azimuthally-symmetric Parker spiral. If the solar wind velocity, $v_{sw}$, is constant, the radial element $dr$ does not change as the magnetic-field lines are advected with the solar wind, while the corresponding length element $dx$ along a magnetic field line changes according to Eq.~(\ref{frozen1}).}
\label{fig:spiral}
\end{figure}

Since the lines are frozen into the radial solar-wind flow whose velocity, $v_{sw}$, is nearly constant, the radial displacement $dr$ corresponding to the field-line element $dx$ does not change as this element is advected with the flow, see Fig.~(\ref{fig:spiral}). We therefore have $dx_0=dr_0\sqrt{2}=dr\sqrt{2}$. From Eq.~(\ref{frozen}) we therefore have
\begin{eqnarray}
dx=dr\sqrt{2}\left(\frac{B(x)}{n(x)} \right)/\left(\frac{B(x_0)}{n(x_0)} \right),\label{frozen1}
\end{eqnarray}
which, after substitution into Eq.~(\ref{dy}) gives
\begin{eqnarray}
dy=\left(\frac{16\sqrt{2}\pi e^4\Lambda \beta n(x_0)}{EB(x_0)} \right)\, dr.\label{y}
\end{eqnarray}
Quite remarkably, we derive that the variable $y$ is equal (up to a constant) to the heliospheric distance $r$. Parenthetically, we note that Eq.~(\ref{y}) would change if the solar wind velocity were not assumed to be constant; see a discussion of this effect in the appendix (\ref{appendix}).

Finally, conducting an additional change of variable $\zeta=\sqrt{M}$, we cast Eq.~(\ref{drift-diffusion}) in the form of a standard 2D radial diffusion equation describing $f(E, \zeta, y)$:
\begin{eqnarray}
\frac{\partial f}{\partial y}=\frac{1}{4}\frac{1}{\zeta}\frac{\partial}{\partial \zeta}\zeta \frac{\partial f}{\partial \zeta}.
\end{eqnarray}
This equation can be solved if the distribution of the strahl electrons is known at some initial position $y_{in}$. We assume that this distribution is narrow, that is, concentrated at $v_\|\gg v_\perp$. Then at larger distances $y\gg y_{in}$ it can be approximated by the standard solution of the 2D diffusion equation:
\begin{eqnarray}
f(E,M,y)=\frac{C(E)}{y}\exp\left(-\frac{\zeta^2}{y}\right)=\frac{C(E)}{y}\exp\left(-\frac{M}{y}\right),\label{f_diff}
\end{eqnarray}
where $C(E)$ is an arbitrary function that may be related to the distribution of fast electrons at the base of the solar wind. This supports similar conclusions drawn by, e.g.,~\cite{smith12}, about the coronal origins of the strahl electrons.  We need not relate function $C(E)$ to the thermal distribution of the core electrons at a given distance, a priori, since the strahl is not in thermal equilibrium with them.  

In the Parker-spiral model, the magnetic field strength changes with the heliospheric distance as $B(r)=B(r_0)(r_0/r)\sqrt{1+{r_0^2}/{r^2}}/\sqrt{2}$, where $r_0$ is the heliospheric distance corresponding to the field-line position $x_0$, as described by Eq.~(\ref{r_0_eq}). We can now re-write the obtained solution (\ref{f_diff}) for the electron-strahl distribution function (again assuming $E\approx v^2$) using the variables $v$, $\mu$, $r$:
\begin{eqnarray}
\label{strahl_model_eq}
f(v,\mu,r)=\frac{C(v^2)}{r}\exp\left\{-\frac{v^4(1-\mu^2)}{\sqrt{1+r_0^2/r^2}}\left(\frac{m_e^2}{16\pi n_0r_0 e^4 \Lambda \beta}\right) \right\},
\end{eqnarray} 
where we have denoted $n_0=n(r_0)$. This completes our solution for the strahl component of the electron distribution function. Except for the undetermined isotropic velocity function $C(v^2)$, this solution does not contain free parameters.  

The width of the obtained strahl distribution function at a given energy can be found directly from this solution. From the exponential factor of Eq.~(\ref{strahl_model_eq}), we find the so-called strahl full width at half maximum, $\theta_{FWHM}$:\footnote{The full width at half maximum is twice as large as the corresponding half-maximum pitch angle $\theta$.} 
\begin{eqnarray}\label{theta_fwhm_eq}
	\theta_{FWHM}= 2 \sin^{-1}\left\{ \frac{16\pi n_0r_0 e^4 \Lambda \beta \sqrt{1+r_0^2/r^2} \ln(2) }{m_e^2 v^4} \right\}^{1/2}.
\end{eqnarray}
Expressions~(\ref{strahl_model_eq}) and~(\ref{theta_fwhm_eq}) are the main predictions of our theory for the electron strahl. 

A simpler expression can be derived using the small angle approximation $\sin^{-1} \theta \approx \theta$. By assuming the typical values for the parameters\footnote{The Coulomb logarithm is estimated as $\Lambda\approx 24-\ln(n^{1/2}/T)$, where $n$~(${\rm cm}^{-3}$) is the density of the particles and $T$~(eV) is their temperature \citep{nrl}. For the estimate,  one needs to consider the particles that thermal velocity is larger than the relative velocity between the scattered (strahl) and scattering (core) particles. We, therefore, substitute here the temperature of the strahl and halo particles, $T\sim 100\,\,{\rm eV}$, and their combined density $n\sim 0.5\,\,{\rm cm}^{-3}$, which is about $5-10\%$ of the core density \cite[e.g.,][]{stverak09}.}  $\Lambda\approx 30$, $\beta\approx 1.05$, and $r_0\approx 1\,\,{\rm AU}$, Eq.~(\ref{theta_fwhm_eq}) can be approximated as:
\begin{eqnarray}\label{theta_fwhm_eq_approx}
	\theta_{FWHM}\approx 24^\circ \left(\frac{{ K}}{100\,\, {\rm eV}}\right)^{-1}\left(\frac{n_0}{5\,\, {\rm cm}^{-3}} \right)^{1/2} \left( 1+\frac{r_0^2}{r^2}\right)^{1/4},
\end{eqnarray}
where $K\equiv m_e v^2/2$.  As previously shown in \cite{horaites18a}, the strahl width varies as the square root of the density, and varies inversely with the energy.

Two important observations should be made about this solution. First, the width of the electron strahl is independent of the overall strength of the magnetic field, as e.g., the term $B(r_0)$ is absent from Eq.~(\ref{theta_fwhm_eq}). The width only depends on the way the magnetic field varies with distance in the Parker spiral. Second, at lower heliospheric distances, $r^2\ll r_0^2$ the focusing effects dominate and the width of the strahl decreases with the distance. At higher distances, $r^2\gg r_0^2$, however, the strahl width {\em saturates} and becomes independent of distance.  

The saturation of the strahl width may seem counterintuitive if one considers that at $r^2\gg r_0^2$ the magnetic-field strength still declines rather rapidly with the heliospheric distance, with the scaling $B(r)\propto 1/r$, and the magnetic focusing effects may be expected to dominate in a nearly collisionless plasma. The resolution to this paradox is that for $r^2\gg r_0^2$ the magnetic field lines are nearly azimuthal in the Parker spiral, in which case the magnetic-field strength declines rather slowly {\em along} the magnetic field line. An electron following a magnetic field line has to travel an increasingly large distance along a rather slowly declining magnetic field, before  considerable focusing can take place. This enhances the effects of collisional broadening relative to the effects of magnetic focusing, leading to the establishment of a universal strahl width in the regime $r^2 \gg r_0^2$, as seen from Eq.~(\ref{theta_fwhm_eq}). 

Our strahl model takes into account the defocusing effects caused only by electron Coulomb collisions. Our results thus present a lower boundary on the width of the strahl. They are however in good agreement with the set of observational data at $1$~AU that we analyze in the next section.  We show that our formula~(\ref{theta_fwhm_eq}) underestimates the width of the strahl in those measurements by only about~$15-20\%$, which indicates that Coulomb collisions provide a dominant contribution to the strahl broadening. In practice, the strahl electrons may also be scattered by plasma turbulence that is ubiquitous in the solar wind, which could further enhance the strahl broadening.

\section{Comparison with observations}\label{obs_sec}
Here we present observations of the angular breadth of the strahl as measured at $r=$1 AU, and compare those observations with the prediction given by Eq.~(\ref{theta_fwhm_eq}). Our measurements of the strahl angular breadth are derived from SWE strahl detector data \citep{ogilvie95}. This observational data set, and the methods used to isolate the strahl population of the eVDF and compute its width, were described in detail in Sections 3.2-3.3, \cite{horaites18a}. We will only provide a brief summary of that analysis here.

The SWE strahl detector was an electrostatic analyzer onboard the Wind satellite. The detector sampled the electron distribution function over a $14\times 12$ angular grid, with a field of view spanning 50$^\circ$ in azimuth and 60$^\circ$ in altitude. The detector produced angular distributions, each measured at a single energy $K$ that was specified with experimental error $\Delta K/K \approx 0.03$. The detector continuously swept through 32 energies ranging between 19.34 and 1238 eV, switching to a new energy every few seconds.\footnote{The eVDFs measured by the SWE strahl detector have recently been made available for download via NASA/GSFC's Space Physics Data Facility's CDAWeb service.} 

\begin{figure}
\hskip-5mm\includegraphics[width=1.05\linewidth]{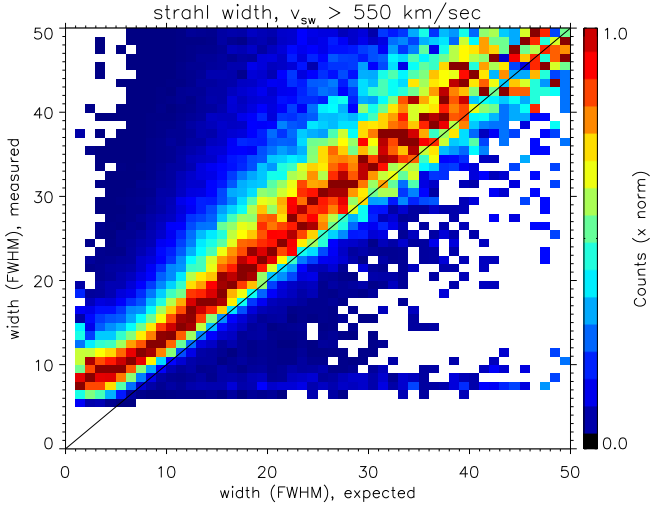}
\caption{The electron-strahl width measured in the fast solar wind intervals (y-axis), as described in \protect\cite{horaites18a}, compared with the analytic prediction (x-axis) given by 
Eq.~(\ref{theta_fwhm_eq}). Here we present the joint probability distribution of the ``measured'' and ``expected'' values of $\theta_{FWHM}$, normalizing each column of the distribution by that column's peak value. The most probable observed widths nearly agree with the predicted values, to within 15-20\%---the data would agree exactly if it fell on the solid diagonal line, shown for reference.  As mentioned in the text, $\theta_{FWHM}$ can only be resolved to a minimum of about $\sim$5--10$^\circ$, which helps explain the relatively large deviation seen between theory and experiment at very small widths.}
\label{strahl_width_comp_fig}
\end{figure}

For each mono-energetic eVDF, a cleaning procedure was applied to separate the strahl from the background halo population. The resulting 2D strahl distribution was then transformed into a pitch-angle distribution, $f(\mu)$, and then fit to a model function:
\begin{eqnarray}\label{fit_eq}
\ln f(\mu) = m (1-\mu) + {\cal Z},
\end{eqnarray}
where the parameters $m$ and ${\cal Z}$ are determined by the fit.\footnote{Technically, we applied a nonlinear fitting procedure that also specified the direction of the magnetic field $\hat B$.  See \cite{horaites18a} for details.} Our fit function, Eq.~(\ref{fit_eq}), is consistent with our model, Eq.~(\ref{strahl_model_eq}). This is seen by noting that each strahl measurement was made at a fixed energy $K = $const., and that the strahl exists in the regime $\mu \approx 1$, where the approximation $(1-\mu^2)\approx 2 (1-\mu)$ can be applied. The strahl width $\theta_{FWHM}$, then follows immediately from Eq.~(\ref{fit_eq}) according to the formula:
\begin{eqnarray}
\label{theta_fwhm_meas_eq}
\theta_{FWHM} = 2 \cos^{-1} \left\{1 + \ln(2)/m \right\}.
\end{eqnarray}

\noindent Note that due to the resolution limit of the strahl detector (4--5$^\circ$ per angular eVDF bin), the angular width $\theta_{FWHM}$ can only be reliably measured to a minimum of $\sim$5--10$^\circ$.

As in \cite{horaites18a}, we restrict our analysis to fast wind intervals, defined as the intervals where the solar wind speed $v_{sw}$ is greater than 550~km/sec. The fast wind tends to exhibit a more pronounced strahl population than observed in the slow wind \cite[e.g., ][]{ogilvie00}, so the strahl properties are, therefore, less subject to error introduced by signal-to-background noise. Each strahl eVDF was measured at a fixed energy $K$, and we only retain distributions for analysis if $K$ was greater than 5 times the core thermal energy, as measured by SWE. We also require the magnetic field direction $\hat B$, as measured by Wind's MFI instrument \citep{lepping95}, to fall within the strahl detector's limited field of view. Other basic selection criteria were applied, such as requiring a minimum number of data points before conducting a fit, and ignoring outlier fits with exceedingly large chi-squared values (reduced chi-squared $>10$). Before applying these criteria, we considered all data measured by the SWE strahl detector between January 1, 1995 and May 30, 2001.

The strahl widths measured by this procedure are compared with the analytical prediction given by Eq.~(\ref{theta_fwhm_eq}), in Fig.~(\ref{strahl_width_comp_fig}). The (peak-normalized) joint probability distribution shown, which compares the ``expected'' and ``measured'' $\theta_{FWHM}$, is comprised of 100,000 width measurements of the fast wind strahl. To calculate the ``expected'' $\theta_{FWHM}$ from the data, we must extrapolate to find the values of $n_0$ and $r_0$ from the local parameters. The value of $r_0$ is calculated according to Eq.~(\ref{r_0_eq}), using the solar wind speed $v_{sw}$ as derived from the proton bulk speed measured by Wind/SWE, and assuming $\omega_S = 2\pi/24.47$ ${\rm days}^{-1}$ \citep{snodgrassulrich90}. We assume that the density $n$ varies with heliocentric distance as $n(r) \propto r^{-2}$, consistent with a constant-speeed solar wind, so that the density $n_0=n(r_0)$ can be extrapolated straightforwardly from the local density measured by Wind. Assuming as before that the alpha particle density is 5\% of the proton density $n_p$, we find that $n_0$ can be estimated by the formula:
\begin{eqnarray}\label{n_0_eq}
n_0 = 1.1 n_p \left(\frac{r_0}{1 \,{\rm AU}}\right)^{-2}.
\end{eqnarray}
Empirically, the value $n_p$ in Eq.~(\ref{n_0_eq}) is the local proton density as observed by SWE's Faraday cup at $r=$1 AU.

We see that our analytic formula~(\ref{theta_fwhm_eq}) shows a reasonably good agreement with the observed strahl broadening, although it slightly underestimates the width, by about 15--20\%. There may be several sources for the systematic error in our derivation. The discrepancy may result from the approximations that we used when we simplified Eqs.~(\ref{coll_op_eq}) and~(\ref{diffusion}), from our evaluation of the parameter $y$ in Eq.~(\ref{y}) where we assumed that the solar wind speed is constant (see appendix \ref{appendix}), or from our idealized assumptions about the Parker spiral that do not take into account large-scale magnetic and density fluctuations (e.g., due to large-amplitude Alfv\'en waves, corotating interaction regions, magnetic discontinuities, shocks, etc.). Importantly, however, the discrepancy may also result from the fact that Coulomb collisions may not be the only mechanism that provides the electron pitch-angle scattering. In the latter case, wave-particle interactions with ambient turbulence may possibly explain extra strahl broadening. Our analysis indicates that Coulomb collisions provide the primary scattering mechanism, however, as the strahl widths would be fully accounted for by an artificial increase of only 30-40\% in the diffusion coefficient (noting that $\theta_{FWHM}$ scales as a square root of the diffusion coefficient).

\section{The electron halo}\label{halo_sec}
The halo population of the eVDF is formed from electrons with high energies, $K\gtrsim 50\,\,{\rm eV}$. Although it shares a similar energy regime with the strahl, the halo population is partly composed of electrons traveling in the {\em sunward} direction. In order to understand the origin of the near-isotropic halo distribution, it is therefore necessary to understand where these highly energetic electrons, which travel large distances without collisions, originate. In our view, the halo may be composed from some of the electrons that run away from the sun as an electron strahl, but later end up on magnetic field lines leading them back to the sun. The halo electrons are, therefore, not produced locally, rather, they are the electrons trapped by magnetic field lines on {\em global} heliospheric scales ($\sim 10 - 20$~AU). 

The idea that the halo electrons are defined by plasma organization on a global heliospheric scale has been entertained previously. For instance, in \citet{scudderolbert79} it was suggested that fast electrons can be scattered back by Coulomb collisions at large distances. Our present treatment, however, demonstrates that the classical Coulomb collisions arrest the magnetic focusing of $100\,\,{\rm eV}$ electrons at about $24^\circ$ {\em independently} of the distance, and, therefore, they cannot easily explain   backscattering. Large-angle, non-diffuse Coulomb scattering, on the other hand, is known to be reduced compared to the leading diffusive effect by a small factor $\sim 1/\Lambda$  \cite[e.g.,][]{li_petrasso93}, and it, therefore, should not be relevant either. Strong backscattering may alternatively be provided by interactions of the electron beam with strong plasma turbulence. Our results in the previous section, however, show that at least for the considered set of measurements the strahl width is rather close to the Coulomb prediction, leaving little room for anomalous turbulent broadening.

\begin{figure}
\includegraphics[width=1\linewidth]{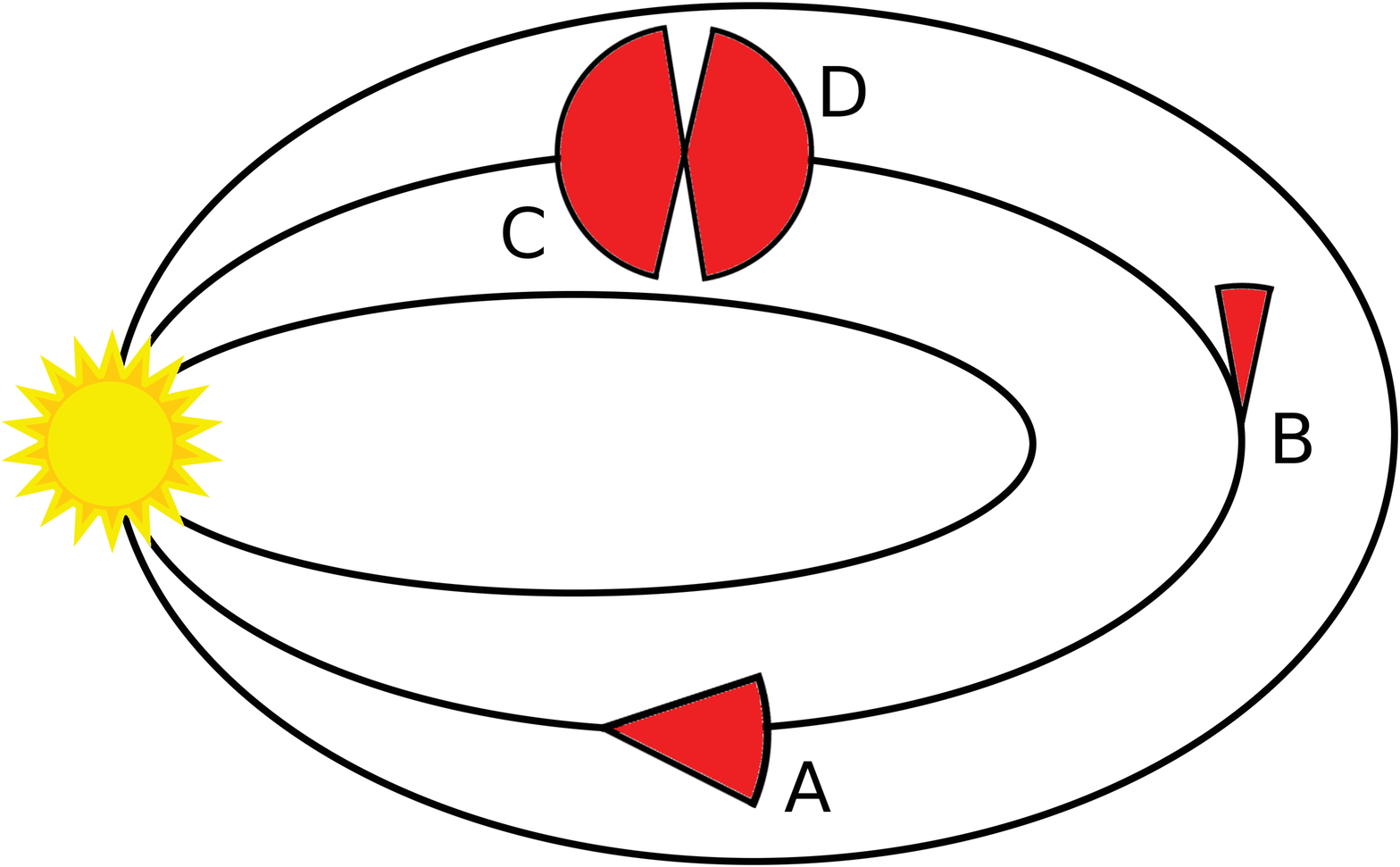}
\caption{Schematic diagram showing how the strahl electrons can evolve, at times labeled (A)-(D), to form a halo population as they travel around a closed magnetic loop. At (A), the electrons have been collimated into a narrow strahl by rapidly diverging magnetic field lines. At (B), the strahl population has been narrowed further as a result of traveling through a weakening magnetic field; however, angular diffusion has made the distribution broader than if the magnetic moment were conserved, thus leading to the {\em universal strahl} given by Eq.~(\ref{theta_fwhm_eq}). At (C), the electrons are broadened by the strengthening magnetic field they experience as they travel towards the sun; although the magnetic field is similar to (A), the distribution is significantly broader here due to the angular diffusion experienced in transit. At (D), the broadened sunward-moving distribution has been reflected to move anti-sunward, due to the magnetic mirror effect. The combined distributions (C) and (D) are observed at the same physical location to form a near-isotropic halo.}
\label{halo_schematic}
\end{figure}

In our present discussion we, therefore, do not specify how exactly the runaway electrons get reflected or redirected to magnetic field lines guiding them back to the sun (a discussion of various possibilities may be found in, e.g., \cite{gosling1993,gosling2001}). We simply assume that such closed trajectories of fast electrons exist. 
As a cartoon example, we may envision closed magnetic-field lines that would be perfect candidates for turning back the fast and nearly collisionless electrons escaping from the sun. This situation is schematically shown in Fig.~(\ref{halo_schematic}).

We now demonstrate that this scenario naturally produces symmetric halo distributions. Consider fast electrons that travel from the sun and form a narrow strahl at point~A in Fig.~(\ref{halo_schematic}). According to our discussion in the previous section, at travel distances exceeding 1~AU the strahl electrons of $100\,\,{\rm eV}$ will have a broadening angle $\theta_{FWHM,\,\,B}\approx 24^\circ$ (Eq.~\ref{theta_fwhm_eq_approx}).  Let us denote the heliospheric distance corresponding to the farthest point of their travel, B, as $r_*$.  When these electrons travel back to the sun, the increasing magnetic field defocuses those electrons, broadening their distribution. Consider the returning electrons ($\mu < 0$) reaching point C whose heliospheric distance is~$r\ll r_*$. The total traveled distance for such electrons is, therefore, $2r_*$, and from Eq.~(\ref{y}) we obtain:
\begin{eqnarray}
y \approx \left(\frac{16\sqrt{2}\pi e^4\Lambda \beta n(x_0)}{EB(x_0)} \right)\, 2r_*.
\end{eqnarray}
Substituting this into Eq.~(\ref{f_diff}), we derive the distribution function of these electrons:
\begin{eqnarray}
f(v,\mu,r)=\frac{C(v^2)}{2 r_*}\exp\left\{-\frac{v^4(1-\mu^2)}{\sqrt{1+r_0^2/r^2}}\left(\frac{m_e^2 \,r}{16\pi n_0r_0 e^4 \Lambda \beta \,2r_*}\right) \right\}.\label{f_strahl_1}
\end{eqnarray} 
From this expression we see that the broadening angle of the returning electrons becomes:
\begin{eqnarray}\label{theta_fwhm_eq_approx_return}
	\theta_{FWHM}= 2 \sin^{-1}\left\{ \frac{16\pi n_0r_0 e^4 \Lambda \beta \sqrt{1+r_0^2/r^2} \ln(2)\, (2r_*/r) }{m_e^2 v^4} \right\}^{1/2}.
\end{eqnarray}

For an estimate, let us assume $r\sim r_0\sim 1\,\,{\rm AU}$, $n_0= 5\,\, {\rm cm}^{-3}$, and $K= 100\,\, {\rm eV}$. 
It is then easy to see that the broadening angle~(\ref{theta_fwhm_eq_approx_return}) becomes $180^\circ$ already for the turning distance of $r_*\approx 8\,\,{\rm AU}$. For even larger turning distances, the broadening angle of the distribution of returning electrons approaches $180^\circ$ before the particles reach point C, so that at point C the distribution is isotropic: $\theta_{FWHM,\,\,C}=180^\circ$. As these electrons get reflected back by the even stronger field they experience closer to the sun, they form an identical ($\mu \rightarrow -\mu$) distribution with width $\theta_{FWHM,D}=\theta_{FWHM,C}$. 
We thus see that just one bounce of highly energetic electrons caught in the magnetic bottle formed by a closed global-scale loop is enough to generate a symmetric halo distribution.\footnote{Strictly speaking, the derivation of our formula~(\ref{f_strahl_1}) is valid only for $1-\mu^2 \ll 1$. We, however, may extrapolate it qualitatively to $1-\mu^2\sim 1$, as at the point where our approach breaks down, the distribution already becomes effectively isotropic.} 

We also point out that our explanation for the halo distribution is not sensitive to the particular origin of the fast sunward-propagating electrons. No matter what their source is, and no matter how narrow their initial distribution at large heliospheric distances is, if these electrons have a chance to reach significantly lower heliospheric distances, the magnetic defocusing and Coulomb pitch-angle scattering effects would turn their distribution into a nearly isotropic halo.  

Obviously, if the fast electrons could be trapped for a long time and allowed to bounce many times, they would eventually thermalize and form a Gaussian distribution. However, the pitch-angle scattering of fast electrons by relatively cold electrons and ions of the core distribution is much more efficient than the electron energy exchange. The fast electrons, though they have sufficient time to isotropize, cannot thermalize before the guiding field lines are advected out by the expanding solar wind.

In our calculations, Coulomb collisions provided the only source of angular diffusion of the halo and strahl. Our picture does not invoke any other mechanisms of pitch-angle scattering. In practice, however, the effective broadening may be somewhat larger due to electron scattering produced by turbulence.  Such extra scattering will, obviously, be even more favorable for the formation of  isotropic halo distributions.         

\section{Conclusions}\label{conclusions_sec}

We have presented a kinetic theory of electron strahl and electron halo formation in the solar wind at electron energies $10\,{\rm eV}\lesssim K \lesssim 200\,{\rm eV}$. In our theory, the strahl is modeled as a weakly collisional population of runaway electrons, that propagate along the lines of a spatially weakening magnetic field forming the azimuthally symmetric Parker spiral. Similarly to many treatments in the past (e.g., \cite{scudderolbert79,lemonsfeldman83,lie-svendsen97,pierrard1999,pierrard2001,landi12,smith12,pierrard2016}), we based our discussion of strahl broadening on Coulomb collisions.  We find that Coulomb scattering counteracts magnetic focusing effects, thus efficiently limiting the attainable beam collimation. For a beam energy of $\sim 100\,{\rm eV}$ and an ambient plasma density of $\sim 5\,{\rm cm}^{-3}$, we estimate that the strahl will not typically be narrower than about $24^\circ$ in the outer heliosphere. If an electron beam collimated this way can be directed back to the sun (say, by following a closed magnetic field line) its distribution will become efficiently isotropized by the combined effects of collisions and magnetic defocusing. In our model, the electron halo population is thus not produced in situ, but rather it originates from runaway strahl electrons trapped in a magnetic field on global heliospheric scales~($\sim 10 -20$~AU). As energetic particles are isotropized by pitch-angle scattering much faster than they get thermalized, the resulting halo distribution is isotropic but not necessarily Gaussian. The particular shape of the halo energy distribution, $C(v^2)$, is not predicted by our model; rather it can be a signature of the particle heating processes operating in the base of the solar wind. 

 
We have demonstrated that our theory of strahl formation agrees reasonably well with observations at $1$~AU, down to the broadening angle of about $10^\circ$ that approximately corresponds to the electron energies $K\sim 200$--$300\,\,{\rm eV}$. The available set of data does not allow us to address smaller broadening angles (due to limits on angular resolution) and, correspondingly, higher electron energies. Our analytic result~(\ref{theta_fwhm_eq_approx}) would, however, predict rather narrow collimation angles of the order of $\theta_{FWHM}\sim 2.4^\circ$ for the high energies $K\sim 1\,\,{\rm keV}$, indicating a rather weak Coulomb broadening. 

The angular width of the returning population, which forms a halo at 1 AU, is similarly predicted to be narrower at higher energies. For a given returning distance, e.g., $r_*\sim 20$ AU, Eq.~(\ref{theta_fwhm_eq_approx_return}) shows that above some finite energy $K\gtrsim 200\,\,{\rm eV}$ the halo distribution at 1 AU will become appreciably narrow ($\theta_{FWHM}<180^\circ$). We therefore see that very energetic electrons may require some mechanism of non-Coulomb (anomalous) scattering in order to produce an isotropic halo population. 

In our observational analysis, we focused on explaining the angular width of the strahl as observed at 1 AU. The SWE strahl detector specialized in sampling high-resolution eVDFs at 1 AU, so our data were particularly suited to examining the angular width. A more comprehensive analysis is required to compare other strahl properties with our model. For instance, the variation of the strahl amplitude with distance and its energy variation $C(v^2)$, remain to be addressed in future studies. We note also that in our analysis, the expressions for the halo and strahl are both derived from the same equation (\ref{f_diff}), so that the appearance of the same function $C(v^2)$ in Eqs.~(\ref{strahl_model_eq})~and~(\ref{f_strahl_1}) may help explain the well-known observation \citep[see e.g.,][]{stverak09} that these populations have similar energy profiles.

Our strahl model is rooted in basic physical phenomena that are known to be relevant in the heliosphere.  Assuming that strahl particles travel along a Parker spiral field, and assuming the diffusion of the distribution is provided by Coulomb collisions alone, we have derived an analytic expression for the strahl distribution. Our formula for the strahl width, Eq.~(\ref{theta_fwhm_eq}), contains no free parameters and thus serves as a useful basis with respect to which other theories of strahl diffusion may be compared. Our observational analysis shows that indeed, Coulomb scattering can almost fully account for the observed strahl width at 1 AU. We developed a halo model to see how far this simple picture, based only on Coulomb collisions, can be carried. Although the halo model presented here is more speculative---for instance because we have not specified the precise mechanism that allows anti-sunward streaming particles to head back towards the sun---it is encouraging to see that the spatial evolution of the magnetic field could conceivably account for the isotropy of the halo population. We believe that the present work may spur future progress in explaining the properties of both suprathermal populations, the strahl and halo, in a unified physical model.

\appendix

\section{Generalizing to non-constant wind speed profiles}\label{appendix}
      
As mentioned in section (\ref{strahl_sec}), our theory assumes that the solar wind speed is constant. Accounting for the finite solar wind acceleration in our model would alter our analysis, notably by stretching out the Parker spiral arms so that strahl particles would travel a longer curvilinear distance before reaching 1 AU. We here estimate the error associated with our assumption of a constant solar wind speed. Let us consider a solar wind with a given non-constant radial speed profile $v_{sw}(r)$.  In our discussion in section (\ref{strahl_sec}), Eq.~(\ref{y}) would then generalize to the form:
\begin{eqnarray}\label{y_precise}
d{\tilde y}=\left(\frac{16\sqrt{2}\pi e^4\Lambda \beta n(x_0)}{EB(x_0)} \right)\frac{v_{sw}(r_0)}{v_{sw}(r)}\,dr.
\end{eqnarray}
Here the variable ${\tilde y}$ is proportional to the travel time. In order to estimate the possible correction provided by the expression~(\ref{y_precise}), we evaluate both $y$ and its more precise value ${\tilde y}$ at $r=1$~AU. We assume that the solar-wind velocity profile is approximated by the fitting expression suggested by \cite{koehnlein96},
\begin{eqnarray}\label{koehnlein_eq}
v_{sw}(r)=v_{sw}(r_0) \exp\left\{-\left(\frac{0.026}{r}\right)^{0.797} \right\},
\end{eqnarray}
where $v_{sw}(r_0)\approx 448\,\,{\rm km/s}$, and $r$ is measured in~AU. We note that the observational data used to fit the profile (\ref{koehnlein_eq}) contains both fast and slow wind intervals (unlike our data analysis which only includes the fast wind), so we may only apply this result as a rough estimate of the speed profile. If we further assume that the collimation process starts at about $r\approx 0.01\,\,{\rm AU}$, then at the distance of $r= 1\,\,{\rm AU}$ we get:
\begin{eqnarray}
{\tilde y}=y\int\limits_{0.01}^{1} \frac{v_{sw}(r_0)}{v_{sw}(r)}dr\approx 1.12 y,
\end{eqnarray}
which means that the more precise calculation would increase the value of $y$ by about~$12\%$.

As the variable $y$ appears in our solution (\ref{f_diff}) to the diffusion equation---the form of which would remain the same under this generalization, after replacing all instances of $y$ with ${\tilde y}$---we see that this more precise determination of the variable $y$ would increase our prediction for the strahl width. Specifically, since $y$ appears in the denominator of the exponential function in Eq.~(\ref{f_diff}), we see that increasing $y$ by $12\%$ would increase our prediction for $\theta_{FWHM}$ (at a given energy) by about 6\%. This correction associated with allowing for a non-constant function $v_{sw}(r)$, though small, would help reduce the 15-20\% gap between our theory and the observations presented in Fig.~(\ref{strahl_width_comp_fig}).

{\em Acknowledgments}---The work of KH and SB was supported by the NSF under the grant no. NSF PHY-1707272 and by NASA under the grant no. NASA 80NSSC18K0646. SB was also partly supported by the DOE grant No. DE-SC0018266, and by the Vilas Associates Award from the University of Wisconsin - Madison. MM acknowledges the DOE grant No. DE-SC0016368 and the DOE EPSCOR grant No. DE-SC0019474.

\bibliographystyle{mnras} 
\bibliography{halo_paper_refs}{}

\end{document}